\documentclass[9pt,twocolumn,twoside]{article}
\usepackage{fancyhdr}
\usepackage{authblk}
\usepackage{abstract}

\renewcommand{\abstractname}{}

\usepackage{graphicx}
\usepackage{textcomp}
\usepackage[margin=1in]{geometry}
\usepackage{enumitem}
\usepackage{epstopdf}
\usepackage{gensymb}
\usepackage{amssymb,amsmath}
\usepackage{xcolor}

\title{\textbf{Snapshot fiber spectral imaging using speckle correlations and compressive sensing}}

\author[1,*]{Rebecca French}
\author[2]{Sylvain Gigan}
\author[1,**]{Otto L. Muskens}

\affil[1]{\small{\textit{Physics and Astronomy, Faculty of Engineering and Physical Sciences, University Road, University of Southampton, Southampton, SO17 1BJ, UK}}}
\affil[2]{\small{\textit{Laboratoire Kastler Brossel, ENS-PSL Research University,CNRS, UPMC-Sorbonne Universit\'{e}s, Coll\`{e}ge de France, 24 rue Lhomond, 75005 Paris, France}}}

\affil[*]{R.French@soton.ac.uk}
\affil[**]{O.Muskens@soton.ac.uk}

\begin{document}
	\twocolumn[
	\begin{@twocolumnfalse}
		\date{}
		
	\maketitle
	
	\begin{abstract}
		Snapshot spectral imaging is rapidly gaining interest for remote sensing applications. Acquiring spatial and spectral data within one image promotes fast measurement times, and reduces the need for stabilized scanning imaging systems. Many current snapshot technologies, which rely on gratings or prisms to characterize wavelength information, are difficult to reduce in size for portable hyperspectral imaging. Here, we show that a multicore multimode fiber can be used as a compact spectral imager with sub-nanometer resolution, by encoding spectral information within a monochrome CMOS camera. We characterize wavelength-dependent speckle patterns for up to 3000 fiber cores over a broad wavelength range. A clustering algorithm is employed in combination with l$_{1}$-minimization to limit data collection at the acquisition stage for the reconstruction of spectral images that are sparse in the wavelength domain. We also show that in the non-compressive regime these techniques are able to accurately reconstruct broadband information.
		\\	
		\\	
		\footnotesize{\textcopyright}  \small   2018 Optical Society of America. Users may use, reuse, and build upon the article, or use the article for text or data mining, so long as such uses are for non-commercial purposes and appropriate attribution is maintained. All other rights are reserved.
\vspace{1cm}
\renewcommand{\abstractname}{\vspace{-\baselineskip}}

\end{abstract}
	
\end{@twocolumnfalse}
]

	\section{\label{sec:level1}Introduction}
	
	Hyperspectral and multispectral imaging are of great importance for acquiring both spatial and spectral information, with applications in environmental sensing to threat detection. For many years, scanning-based hyperspectral imaging techniques, such as pushbroom and whiskbroom spectral imaging systems, have been at the forefront of sensing approaches. While they have been widely used in remote sensing applications, scanning systems have many drawbacks, such as the requirement of performing many sequential measurements in order to reconstruct an image.  A more desirable approach for many scenarios is the use of snapshot technologies, which acquire both spatial and spectral information in one measurement. Commonly seen in astronomy, Integral Field Spectrometers (IFSs) aim to acquire full hyperspectral data cubes within single snapshot images, with many based on lenslet arrays, fiber bundles, and slicing mirrors \cite{Hagen2013,Bacon1988,Peters2012,Dwight2017, Drory2015,Eisenhauer2003}. However, most devices rely on grating-based wavelength characterization, with device footprints scaling with spectral resolution. Alternative techniques using thin-film filters have also been employed which can be scaled more easily for portable applications. However, in many cases, these are limited to narrow spectral ranges with a restricted spectral resolution. In recent years, a new category of spectrometers has been proposed, which exploits the natural properties of complex materials \cite{Kohlgraf2010,Xu2010, Redding2012,Chakrabarti2015,Liew2016,Redding2013b, Mazilu2014,Hang2010, Valley2016, French2017}. The use of random media in controlling information has followed our understanding that disorder provides new opportunities for highly multi-mode data processing \cite{GiganRotter2017,Popoff2010, Liutkus2014, Vellekoop2007,Katz2012,Andreoli2015}. Complex systems have the advantage of allowing transmission over broad wavelength ranges with a spectral resolution that depends mainly on the scrambling strength of the medium. Recently, other solutions for spectral imaging have been demonstrated based on thin diffractive optical elements \cite{Wang2015, WangJOSA18} and by exploiting the memory effect of a ground-glass diffuser \cite{Sahoo2017}.
	
	Next to complex scattering media, multimode fibers are receiving considerable interest for imaging and spectroscopy owing to their high throughput and their use in applications such as remote sensing and endoscopy \cite{Ploschner2015,Porat2016}. By exploiting wavelength-dependent speckle patterns from a multimode fiber, a spectral resolution of picometers in the near-infrared and nanometers in the visible region has been demonstrated \cite{Redding2014,Wan2015}. By taking advantage of off-the-shelf fiber technology, the potential cost of these speckle-based spectrometers can be significantly lower than traditional spectroscopy devices.

	\section{\label{sec:level2}Fiber imaging spectrometer}
	
	An important aspect of hyperspectral imagers is the generation of vast amounts of data, accompanied by challenges in processing. Modern techniques, such as coded aperture snapshot spectral imaging (CASSI), are moving towards minimizing data collection at the acquisition stage by using computational methods to sample below the Nyquist-Shannon limit \cite{Wagadarikar2008,Gehm2007, Cao2016,Willett2013,IAugust2013,YAugust2013,Rueda2015,Fowler2014,Lin2014}. If a signal is sparse, with only few non-zero terms, a lot of redundant information is collected to meet the requirements of the sampling theorem. It was shown by Cand\`es, Romberg, Tao, and Donoho that a signal could be sampled below this classical limit and fully reconstructed by minimizing the number of non-zero components in the solution \cite{Candes2006b,Donoho2006}. This led to a new era of data collection, known as compressive sensing (CS). This emerging field has enabled a new wave of imaging and sensing modalities, such as single-pixel cameras \cite{duarte2008single}. Recently, it has been shown that multiple scattering of light works in harmony with CS, as information from all inputs can be accessed from a small number of measurements, due to mixing of input and output degrees of freedom \cite{Liutkus2014}. In previous work, we used a multiple scattering layer of GaP nanowires to demonstrate reconstruction of hyperspectral information on a 9x9 pixel array \cite{French2017}.
	
	Here, we use a rigid multicore multimode fiber array (MCMMF) in combination with a monochrome CMOS camera to demonstrate a high resolution, high throughput, imaging spectrometer with over 3000 individual pixels. The spectrometer makes use of the interference between modes within each core of a 30~cm-length fiber bundle to produce frequency-dependent speckle patterns. We take advantage of CS, in combination with a clustering algorithm, to measure a spectral intensity transmission matrix (STM) of each core within the MCMMF, demonstrating efficient acquisition of spatial and spectral information in one measurement for snapshot spectral imaging.
	
\begin{figure*}[t!h]
\centering
\includegraphics[width=0.7\linewidth]{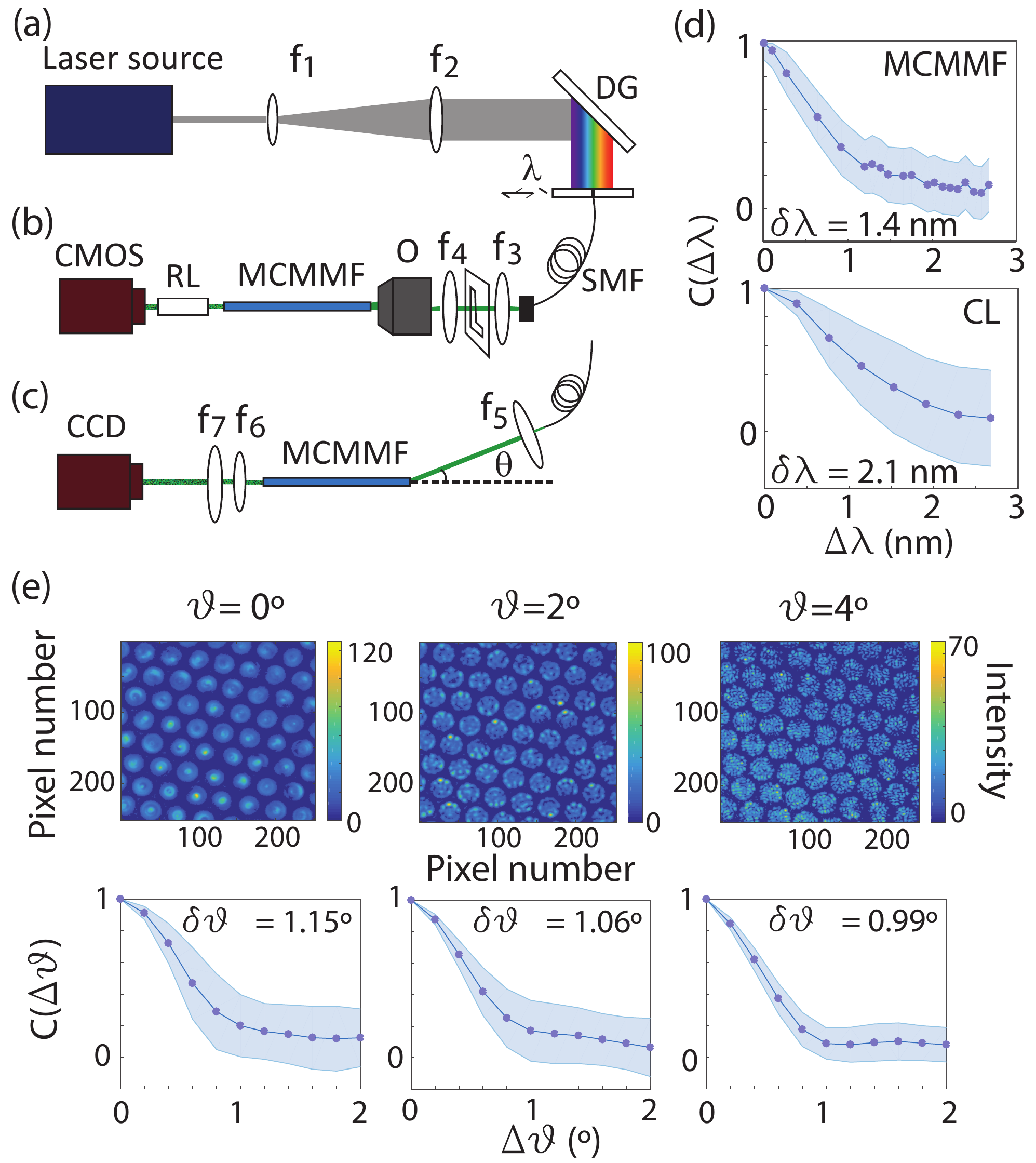}
\caption{Characterizing a fiber imaging spectrometer. (a) Setup showing broadband supercontinuum calibration laser (CL), 10$\times$ beam expansion (f$_{1}$ and f$_{2}$), and diffraction grating (DG). One wavelength from the resulting spectrum is selected using a 10~\textmu m pinhole (PH) and single mode fiber (SMF). (b) Light from the SMF is collimated using lens f$_{3}$. Objects are imaged onto the input of a multicore multimode fiber (MCMMF) using a 3~cm lens (f$_{4}$) and a 10$\times$ objective (O). Light transmitted through the fiber bundle is imaged onto a CMOS camera using a relay lens pair with 3~cm focal length (RL). (c) SMF is mounted on a rotation stage, with lens f$_{5}$, to deliver light at angle, $\theta$ to the MCMMF. Output light is imaged onto a monochrome CCD using f$_{6}$ and f$_{7}$. (d) Spectral correlation width of: the MCMMF, with full-width half maximum (FWHM) of 1.4~nm obtained using narrowband laser source; the system when using the supercontinuum calibration laser source (CL), resulting in FWHM of 2.1~nm. (e) Camera images of speckle patterns produced by MCMMF cores, with incident angles of $\theta$=0\textdegree, 2\textdegree, 4\textdegree. The corresponding angle correlation functions are shown for each $\theta$. The shaded regions in (d) and (e) correspond to the standard deviation across all fiber cores. }
\label{fig:Experiment1}
\end{figure*}
	
	A collimated light source was imaged onto the facet of the MCMMF (Edmund Optics, Fiber optic image conduit, 40-643), as shown in Fig.~\ref{fig:Experiment1}(a) and (b). The resulting beam spot size on the MCMMF facet spanned the total diameter of the fiber. The 308.5~mm length of MCMMF, with outer diameter 3~mm, consisted of approximately 3000 multimode fibers, which defined the number of spatial pixels in the resulting spectral images. Each fiber core had a diameter of 50~\textmu m. The number of modes coupled into each step index fiber using this set up, was approximately 85 modes at a wavelength of 670~nm (see Appendix A for calculation). The end of the MCMMF was imaged onto a monochrome camera using a relay lens consisting of a matched pair of two achromatic doublet lenses, each with focal length of 3~cm. A 12-bit, 5 MPixel monochrome CMOS array (AVT Guppy) of 2.2~\textmu m x 2.2~\textmu m pixel size was used to measure the output speckle information. The size of individual speckle grains was approximately matched to the size of one camera pixel in order to optimize the information content of the image. The absence of a polarizer at the fiber output maximized the throughput of the system to 45\% but reduced the speckle contrast by a factor of $\sqrt{2}$.
	
	In order to probe the spectral correlation bandwidth of the MCMMF, a Continuous Wave External Cavity Diode Laser (ECDL), with central wavelength of 780~nm, was tuned over a bandwidth of 5~nm \cite{Woods2016}. Measurements of consecutive speckle patterns produced by monochromatic light were taken as the wavelength was changed. The spectral correlation between the patterns is presented in Fig.~\ref{fig:Experiment1}(d) and shows a width of 1.4~nm. For the testing and calibration of our imaging spectrometer over a large bandwidth, a spectrally tunable source was obtained by filtering from a supercontinuum laser (Fianium, SC-400-2) using a 600 lines / mm reflection diffraction grating and a 10~\textmu m single-mode optical fiber, resulting in a output spectral bandwidth of 0.5~nm. The input wavelength was confirmed by a commercial spectrometer (Ocean Optics, USB4000). The total correlation width of the fiber spectrometer and source was therefore 2.1~nm (see Fig.~\ref{fig:Experiment1}(d)). As the calibration light had a greater spectral bandwidth than the spectral correlation width of the MCMMF, the speckle contrast was reduced. 
	
	The spectral correlation of multimode fiber spectrometers scales with 1/L, where L is the length of the fiber, implying that high spectral resolutions can be achieved with long MCMMFs. This can be seen in Appendix B, where the correlation widths are shown for two MCMMFs with L~=~25.4~mm and L~=~152.4~mm, respectively. Every multimode fiber within the image conduit produces an independent and unique speckle pattern which changes with wavelength. By storing these wavelength `fingerprints' for every spatial position, the original spectral signals can be determined. 
	
	An important aspect of fiber-based spectral imaging systems to consider is the incident angle with which light enters the fiber system. As angle dependence can affect the calibration of a speckle spectral imaging system, interesting quantities to investigate are the short range correlations which are present in measurements such as, in this case, the angular memory effect. Figure~\ref{fig:Experiment1}(c) shows the experimental setup used to probe the angle dependence of speckle patterns produced by a 30~cm-length of MCMMF with core diameter 50~\textmu m. Speckle patterns were measured after light of a known incident angle was sent through the system, resulting in images such as shown in Fig.~\ref{fig:Experiment1}(e). For an incident angle of 0\textdegree, N$_{m}$ corresponds to roughly a single transverse mode, showing that light can only couple to a small number of modes at small incident angles. With increasing incident angle, a larger number of modes is excited, as deduced from the size and number of speckle grains produced by each core. The corresponding angle correlation functions for $\theta$ = 0\textdegree, 2\textdegree, 4\textdegree, are shown beneath the camera images. With a full-width at half maximum of around 1$^\circ$, the numerical aperture for single speckle imaging amounts to 0.017, or an F-number of 29. For our spectral imaging system, we used an incident angle of around 3.5\textdegree, which optimizes the number and size of speckle grains onto our camera pixels, thereby allowing a high spectral resolution. 
	
	\section{Algorithm for spatial and spectral imaging}
	
\begin{figure*}[t!h]
		\centering
		\includegraphics[width=0.8\linewidth]{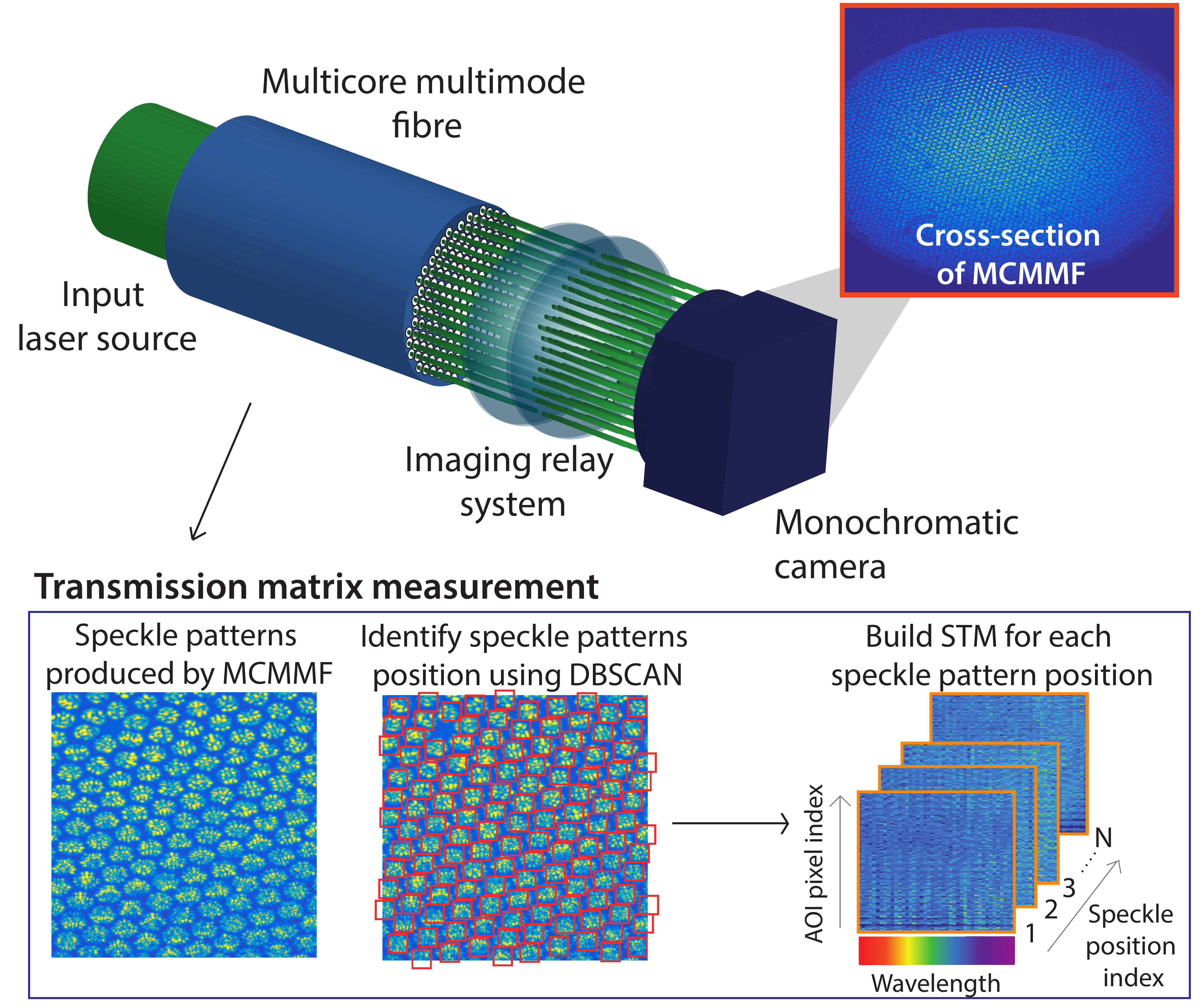}
		\caption{A schematic showing the principle of the MCMMF STM calibration. Light travels through a MCMMF and the cross-section of the end of the fiber is imaged on to a camera. A clustering algorithm, DBSCAN, detects the positions of the speckle patterns produced by each fiber core on the camera. These coordinates are then used to calibrate a STM by measuring wavelength-dependent speckle patterns at every position and storing them in a 3-dimensional data cube. The coordinates are also used to search any arbitrary image for spectral signatures at those positions after calibration. }
		\label{fig:Experiment2}
\end{figure*}

	A spectral intensity transmission matrix (STM) was obtained for every fiber core, as illustrated in Fig.~\ref{fig:Experiment2}. The STM maps spectral information of the input onto a spatial distribution of camera pixels within each fiber core. Each of the camera pixels, therefore, has the ability to reconstruct spectral information using the STM. In order to build the STM, the center positions of every fiber core seen on the camera image must be known. Because of the large number of cores and their slightly irregular placement in the bundle, we employ a Density-Based Spatial Clustering of Applications with Noise (DBSCAN) algorithm to identify the fiber cores from the camera image and to determine their corresponding center position \cite{Ester1996}. This algorithm significantly reduces the time taken to gather the fiber core locations, and results in a relatively low number ($<$1\%) of defects in our reconstructed images where the algorithm failed to identify a core. The zoomed-in region of the MCMMF in Fig.~\ref{fig:Experiment2} shows that, for homogeneous illumination with a single wavelength, some of fibers appear to produce similar patterns, but variations in the speckle patterns are observed between all of the fibers. These variations reflect small differences between individual fibers in the bundle, showing that an individual layer of the STM is required for each core.
	
	To calibrate the STM, the source was scanned across a number of wavelengths, X, and the resulting speckle information was recorded by the 12-bit monochromatic camera. An area of interest (AOI) consisting of a number of camera pixels, Y, was selected for each core. The pixel information collected was unwrapped into a column vector of length Y, and was stored in the STM for every spectral channel. An example STM for a single core is shown in Fig.~\ref{fig:Experiment2}.
	
	Our imaging spectrometer is described by a linear equation y~$=$~Ax~$+$~c, where y is the vector of selected camera pixels, A is the STM, x is the spectral input, and c is inherent noise in the system. Since A is known and y can be measured with our camera, we can solve the problem for x using a matrix inversion. However, we need to employ a computational technique that can compensate for the noise, c.
	
	One approach utilizes the properties of CS to suppress background noise in the reconstructions. To take advantage of CS, the domain in which the measurement is sparse should be known. For our imaging spectrometer we assume that the spectra in each spatial channel are sparse. As sufficient spectral information is contained within a small area of the speckle patterns, the area of interest collected at the acquisition stage can be minimized. In general, if fewer camera pixels are stored than calibrated spectral bands, the system is under-determined and cannot be solved using classical inversion techniques. In this case, we can exploit the realm of CS to find a solution. 
	
\begin{figure*}[t!h]
		\centering
		\includegraphics[width=\linewidth]{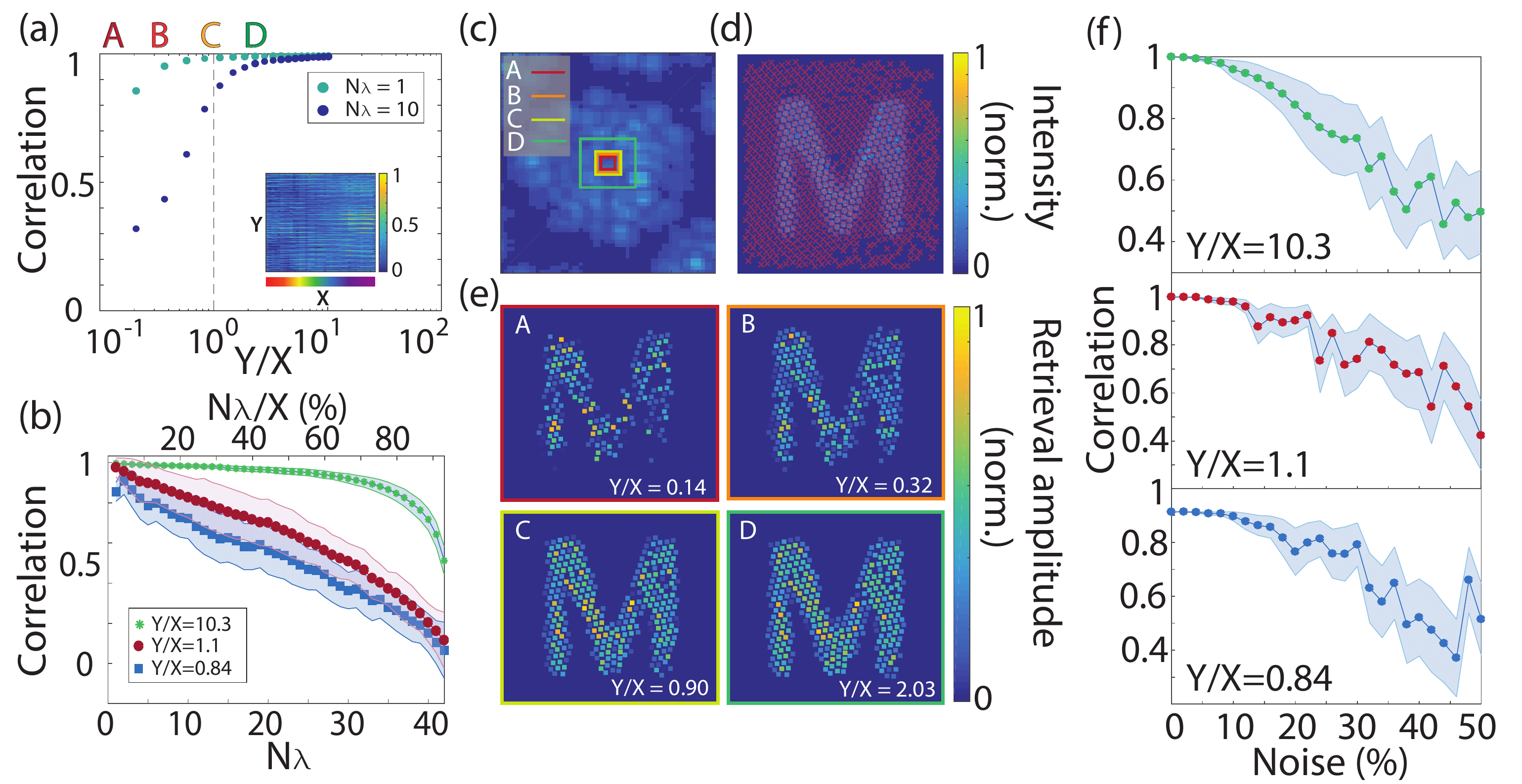}
		\caption{Imaging spectrometer robustness. (a) A STM illustrates the ratio between the number of camera pixels sampled, Y, and the number of calibrated wavelength increments, X. The resulting correlation between a known input spectrum and the resulting output spectrum is determined as Y/X is varied for two different spectra: N$_{\lambda}$=1 and N$_{\lambda}$=10, where N$_{\lambda}$ is the number of wavelengths contained within the spectrum. The letters A, B, C, and D correspond to those labeled in (c) and (e). (b) Correlation between a known input spectrum and reconstructed spectrum, with an increasing N$_{\lambda}$, for three different sampling rates. The upper x axis shows the percentage of wavelengths in the spectrum from the STM wavelength calibration. (c) A speckle pattern produced by one fiber core. Illustrated are four different areas selected to build a STM, each labeled A to D enclosing the smallest to largest pixel number, respectively. (d) Camera image showing resulting light after light travels through the multimode multicore fiber. The coordinates identified by the DBSCAN clustering algorithm are plotted to show the areas where the STM will be applied to. (e) Four reconstructed images, each with a different sampling rate (Y/X=0.14, 0.32, 0.90, 2.01) for both the STM and the output data recorded. (f) Computational experiment to probe the robustness of the CS technique for three different sampling rates as artificial random i.i.d. noise is increased.}
		\label{fig:Figure3}
\end{figure*}

\begin{figure*}[t!h]
		\centering
		\includegraphics[width=\linewidth]{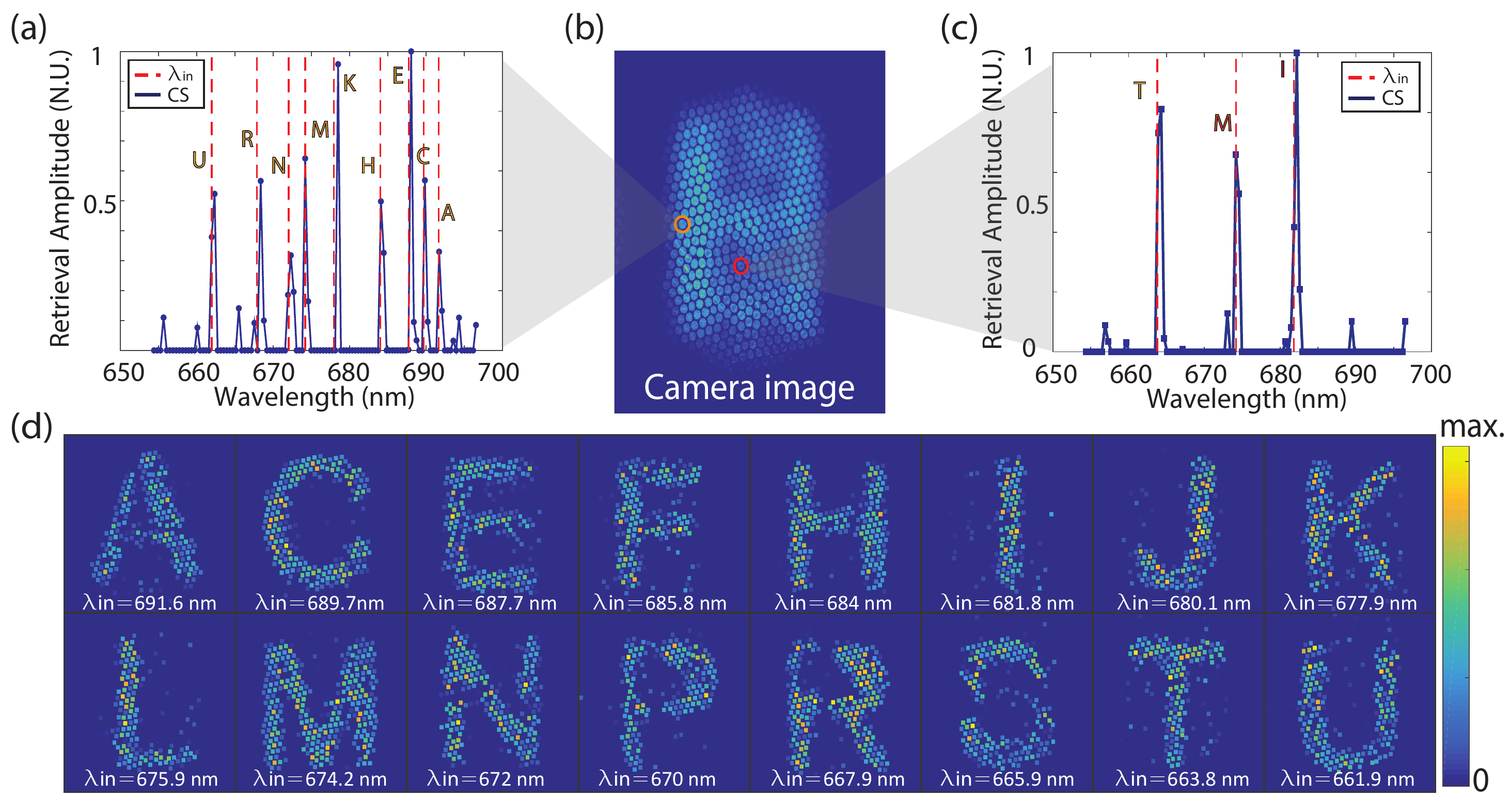}
		\caption{Reconstructing spatial and spectral information. (a) Spectrum from one spatial position (orange circle) reconstructed using the CS technique, with the letters corresponding to the wavelength labeled. The reconstructed spectrum is compared with that measured by a commercial spectrometer. (b) Conjugate camera image of 16 letters, each measured using a different wavelength. (c) As in (a), a spectrum reconstructed from the speckle pattern highlighted with red circle in (b). (d) Reconstructed spatial information for each spectral channel for the speckle image in (b). In total 16 letters are reconstructed from one application of the STM to the output information. }
		\label{fig:Figure4}
	\end{figure*}

	For each of the N spatial coordinates (fiber cores), we computed the following $l_{1}$-norm operation:
	\begin{equation}
	minimize \: \|A_{N}x_{N} - y_{N}\|_{{1}} \: \: \: s.t. \: x \geqslant 0.
	\end{equation}
	This method was implemented using an open-source convex optimization package, CVX \cite{CVX2013}. In this way, spectral information can be obtained separately for each core, and a spectrum can be observed by plotting the resulting intensities against wavelength as determined by the STM. To plot spectral images, spatial intensities at a selected wavelength were assembled using the known fiber coordinates, and a kernel convolution was performed to plot the points with the equivalent size to the fiber core diameters. This procedure enabled the spatial reconstructions shown in Fig.~\ref{fig:Figure3} and Fig.~\ref{fig:Figure4}. 
	
	This technique is able to reconstruct an input signal using a small number of measurements, and therefore enables us to reconstruct a sparse input, x, when sampling below the rate permitted by the Nyquist-Shannon theorem. In the following, we use this method to recover both under-sampled and oversampled speckles for sparse and more broadband signals, respectively.
	
	\section{Characteristics of fiber imaging spectrometer}
	
	For the STM to reconstruct spectral information with high fidelity, it is necessary to select an AOI from the speckle patterns that contains sufficient information. Figure~\ref{fig:Figure3}(a) explores the scaling of the reconstruction quality against the ratio of the number of AOI camera pixels, Y, to the number of independent spectral components, X. The STM was calibrated between 694~nm and 609~nm, and contained 43 independent wavelength channels. The parameter Y$/$X defines the sampling condition, with the Nyquist-Shannon sampling theorem indicated by the dashed line at Y$/$X$=1$, with Y$/$X $> 1$ corresponding to oversampling and Y$/$X $< 1$ to under-sampling. The reconstruction quality is quantified by the correlation between the original spectrum and the reconstructed spectrum. For the STM and the second ``output'' dataset, two different measurement sets were used which were obtained consecutively using the same setup. 
	
	In Fig.~\ref{fig:Figure3}(a) two different conditions were tested, corresponding respectively to spectra consisting of a single wavelength ($N_{\lambda}=1$, relative sparsity ratio 2.5\%), and ten wavelengths ($N_{\lambda}$=10, relative sparsity ratio 25\%), where $N_{\lambda}$ denotes the number of nonzero wavelength components within the spectrum, and the relative sparsity ratio $N_{\lambda}/$X gives the number of nonzero elements as a fraction of the total number of calibrated wavelength bands \cite{Liutkus2014}. Above the sampling limit (Y$/$X $> 1$), the STM is efficient in recovering both spectra. Below this threshold, the spectral reconstruction deteriorates. It is evident that single wavelengths can be reconstructed with a very small amount of information, with the smallest STM containing 9 camera pixels for each calibrated wavelength. In the case of the more dense spectrum with 25\% sparsity, we see a marked decrease in the under-sampling regime. Taking a correlation $>0.5$ as the criterion for reconstruction quality, the smallest STM to provide reasonable reconstruction is obtained for Y$/$X$=0.45$.
	
	The condition Y$/$X$=1$ is equivalent to the sampling condition in our spectral imaging system, and describes the shape of the STM and its mathematical inversion capability. This definition does not take into account additional sample-induced correlations present in the speckle pattern. These correlations effectively reduce the total information contained in the STM and tend to shift the curves in Fig.~\ref{fig:Figure3}(a) to the right, i.e. for a larger effective Y$/$X. In our imaging configuration, a single speckle grain covers around 4 camera pixels.
	
	In Fig.~\ref{fig:Figure3}(b) the same parameter space is explored in more detail against the number of nonzero spectral components N$_\lambda$ (and relative sparsity ratio, top axis), for three different values of the sampling rate Y$/$X. The MCMMF system can reconstruct a spectrum containing nonzero elements of up to all 43 calibrated wavelengths simultaneously when sampling far above the sampling threshold. In particular, we find high correlations above 0.9 for $N_\lambda=30$ (relative sparsity ratio 75\%). When sampling at and below the threshold, for Y$/$X$=$1.1 and 0.84 respectively, the correlation decreases, but reasonable performance (correlation $>0.5$) is obtained up to 35 and 25 wavelengths. Overall the trends show a decrease in reconstruction capability with increasing relative sparsity ratio, as a consequence of exploiting CS. Shaded regions in this and following Figs. indicate the width (standard deviation) of the distribution in correlation amplitudes when evaluating all individual fiber cores. In this case, consistent behavior is shown over the entire bundle, although a greater variation is seen for lower sampling rates. This result shows that oversampling pixel information allows the reconstruction of broadband signals. Furthermore, sparse spectral signals can be reconstructed using our system when sampling below the Nyquist-Shannon limit.
	
	While Figs.~\ref{fig:Figure3}(a) and ~\ref{fig:Figure3}(b) show the performance averaged over all fiber cores, it is of interest to investigate how reconstruction quality affects the imaging capabilities. Figure~\ref{fig:Figure3}(c) shows the zoomed-in detail of a typical speckle pattern produced by one of the cores in the fiber bundle. Four areas are highlighted corresponding to the pixel area of the speckles selected to build the STM, with the smallest to largest boxes denoted using the letters A to D, respectively. For the letter reconstructions presented here and further below, the STM was calibrated with 111 wavelength channels with a separation of 0.4~nm. In Fig.~\ref{fig:Figure3}(d) the raw camera image is shown, with the coordinates identified by the DBSCAN clustering algorithm (red crosses) indicating the areas the STM is applied to. The resulting spatial reconstructions are shown in Fig.~\ref{fig:Figure3}(e), labeled A to D, when Y$/$X=0.14, 0.32, 0.90, 2.03, respectively. The reconstructions show the values of the selected component of x corresponding to the input wavelength. Reconstructions B, C and D share a close resemblance with the camera image which consisted of a single wavelength component. The agreement shows that good imaging performance is obtained when the STM and resulting output are sampled at rates lower than the sampling criterion of Y$/$X$=1$. We see a noticeable degradation in reconstruction A, where Y$/$X$=0.14$, due to insufficient spectral data contained within a very small number of pixels (4 pixel $\times$ 4 pixel area). However, it is still possible to recognize the shape of the letter in the wavelength channel above the noise background. Importantly, other wavelength channels do not show this shape (see supporting information in Appendix D).
	
	A challenge in image reconstruction is posed by non-specific signals or noise that can be present in the scene itself, in the imaging system or in the detection electronics. To probe the robustness of the system against noise, we tested the reconstruction  of a single spectral component in the presence of added computational noise. For this purpose we used the experimental dataset of Fig.~\ref{fig:Figure3}(b) at $N_\lambda=1$. We then computationally added random independent, identically distributed (i.i.d.) noise, of relative strength increasing from 0 to 50\% of the average intensity, to the measured dataset. Fig.~\ref{fig:Figure3}(e) shows the spectral reconstruction capability for the three sampling rates probed in Fig.~\ref{fig:Figure3}(b). We find a correlation $>0.5$ for up to 40\% noise, with most spectral information being reconstructed correctly and without a strong dependence on the sampling rate Y$/$X. This confirms that our spectral imaging system is robust to noise, irrespective of sampling rate.
	
	\section{Hyperspectral reconstruction}
	
	To demonstrate the combined spatial and spectral reconstruction capabilities of the MCMMF system we used a composite image consisting of a superposition of 16 individual, experimentally-measured letter-shaped objects, with each object located at a different optical wavelength. Figures~\ref{fig:Figure4}(a)-(d) summarizes the experiment, where the raw composite image is shown in Fig.~\ref{fig:Figure4}(b). Small variations in intensity over the image are due to different weights of the individual letters, however it is not possible to distinguish the individual components from the composite by eye. 
	
	The CS technique, however, is capable of separating the individual spectral contributions. We extracted spatial and spectral information about the original objects using a sampling rate of Y$/$X$=4$. For this experiment, the STM was calibrated in increments of 0.4~nm over a wavelength range of 654~nm to 697~nm. Spectra reconstructed from two selected fiber cores are shown in Figs.~\ref{fig:Figure4}(a) and \ref{fig:Figure4}(c), for cores marked with the orange and red circles, respectively. The central wavelengths of the known spectral components are illustrated with red dashed lines. As the linewidth of the illumination light source was slightly larger than the spectral band spacing, we observe reconstructions for three consecutive wavelengths around the known wavelength. Despite the core containing a large amount of spectral data, the STM can correctly identify the main wavelengths corresponding to letters. It is evident that there are some smaller reconstruction peaks within the spectrum: some are due to a partially illuminated fiber core with a low intensity at the corresponding wavelengths (e.g. letter S at 665.9~nm), and others correspond to reconstruction noise. The spectra of the two selected cores have relative sparsity ratios of 12\% and 4\% respectively.
	
	Figure~\ref{fig:Figure4}(d) shows the corresponding spatial reconstructions, where all 16 letters are efficiently reconstructed from the snapshot composite image in (b), with very little apparent cross-talk. As discussed, some fiber cores in the bundle are missing from the calibration, giving rise to a few gaps in the total image, and hence a minimal amount of spatial information is lost from the reconstructions. Furthermore, the observed gaps in letters such as `P' and `R' are features inherent in the stencil-type letter aperture used. Small amounts of background noise in the reconstructions account for the low noise values in the spectral reconstructions. However, overall we find that the spectral images across $>1000$ spatial channels can be reconstructed with good accuracy in a single measurement.
	
	\section{Discussion}
	
	Speckle-based spectroscopy using complex media has made much progress in recent years, in an attempt to bridge the gap in current spectroscopy techniques, for high resolution, portable applications. While there have been many breakthroughs using disordered systems, there are also considerations to make when calibrating these speckle sensing devices. As speckle patterns are sensitive to changes in the properties of input light, fiber orientation, and temperature, they can be used as highly accurate sensors \cite{Pan1994, Choi1997, Liu2007, Amitonova2015}. However, if these parameters are not well known, they can cause decorrelation of the calibrated matrix. 
	
	In the case of our spectral imager, the stability of the system was found to be constant throughout the duration of our measurements, on the order of a few hours. Environmental sensitivity derived from temperature changes has been shown to be easily compensated for by observing shifts in wavelength identification, implying that additional data does not need to be collected to account for temperature variations post-calibration \cite{Redding2014}. As the MCMMF has a rigid structure, the resulting speckle patterns are insensitive to accidental bending, which often limited previous demonstrations of multimode fiber-based spectroscopy due to the de-correlation of speckle patterns with fiber bending.
	
	Perhaps one of the main limitations of the MCMMF system is its finite angular correlation width. The spectral imaging system described here had a correlation width of approximately 1$^\circ$, which was proportional to the incident angle of the input light (see Appendix E).
	It is possible, however, to increase the aperture by calibrating speckles for different incident angles, and by understanding how light of different incident angles interacts to form complex speckle patterns. Such an approach could even be used to obtain combined spatial and angular information on the light field, which would allow simultaneous spatial and angular hyperspectral characterization of light fields. Ultimately, the angular range of the MCMMF is limited by the evanescent coupling between adjacent cores, which was found to take place above 4.5$^\circ$ angle of incidence.
	
	Our approach has the advantage of allowing snapshot spectral images, therefore gaining spatial and spectral information instantaneously, rather than relying on slower scanning-based methods. Combined with CS, we can limit the amount of data to be collected at the acquisition stage, further reducing computational bottlenecks. Due to the sparsity condition, working in the compressive regime is only suited to recovering images with a small amount of spectral information compared to the calibrated number of spectral bands. However, the same CS techniques are still able to accurately reconstruct broadband signals when sampling above the Nyquist-Shannon limit, as demonstrated in Figs.~\ref{fig:Figure3}(a) and \ref{fig:Figure3}(b), and Fig.~\ref{fig:Figure4}(a), and only require a relatively small number of pixels ($>$ Nyquist-Shannon limit) to be selected for calibration and detection. The speed and efficiency of the computational processes used here are dependent on the size of X and Y and require on the order of seconds to minutes to compute, depending on the size of the STM. However, the algorithm has not been optimized for speed, and the processing times may be significantly reduced when using more specialized equipment.
	
	With the promising sub-nanometer spectral resolution achievable using a 30~cm-long MCMMF, our snapshot spectral imaging system can resolve fine spectral details for approximately 10$^3$ independent spatial channels. Furthermore, for broadband applications, where a high spectral resolution isn't critical, a shorter MCMMF with lower spectral resolution (as in Appendix D) could be used, to increase the speckle contrast within each wavelength increment. This suggests that fiber-based spectral imaging can be easily tailored to a desired application.
	
	\section{Conclusions}
	
	In conclusion, we have shown that a multicore, multimode fiber bundle can be used as a frequency characterization component in a high-throughput imaging spectrometer for snapshot spatial and spectral measurements with sub-nm wavelength resolution. Utilizing a clustering algorithm we can rapidly identify up to 3000 individual fiber cores for fast calibration of the STM and recovery of spatial information.  Using wavelength- and fiber core-dependent speckle patterns in combination with a transmission matrix approach, we can exploit the realm of compressive sensing to minimize data collection at the acquisition stage for efficient detection and reconstruction of sparse spectral information. Our technique is robust to noise when sampling at rates above and below the limit dictated by the Nyquist-Shannon theorem, for complex and sparse spectra, respectively. The imaging spectrometer and its corresponding computational technique are versatile in their design, and the calibrated wavelength range can be tailored for the application. This approach combines a low-cost, compact fiber component with fast computational techniques for snapshot spectral imaging, removing the need for scanning-based systems.
	
	\section*{Funding}
	Defence Science \& Technology Laboratory (DSTL) (DSTLX1000092237); H2020 European Research Council (ERC) (278025); Engineering and Physical Sciences Research Council (EPSRC) (EP/J016918/1).

	\appendix
	\section{Number of fiber modes}
	
	The number of speckle grains contained within each speckle pattern is approximately equivalent to the number of modes light is coupled to in every fiber. The number of modes in each fiber is calculated using the following equation:
	\begin{equation}
	N_{m} \approx \frac{4}{\pi^2} V^2 = \frac{4}{\pi^2}(\frac{\pi D \times \mathrm{NA}}{\lambda})^2,
	\end{equation}
	where $V$ is the V number of the step-index fibers, $D$ is the diameter of each fiber core, NA is the numerical aperture of the fiber, and $\lambda$ is the propagating wavelength of light \cite{Saleh2001}. For a MCMMF with $D$~=~50~\textmu m, NA~=~0.06, and propagating wavelength $\lambda$~=~670~nm, N$_{m}$ is calculated to be approximately 85 modes, corresponding to the number of speckle grains imaged on the camera.
	
	\section{Spectral correlation bandwidth dependence on fiber length}
\begin{figure*}
		\centering
		\includegraphics[width=0.9\linewidth]{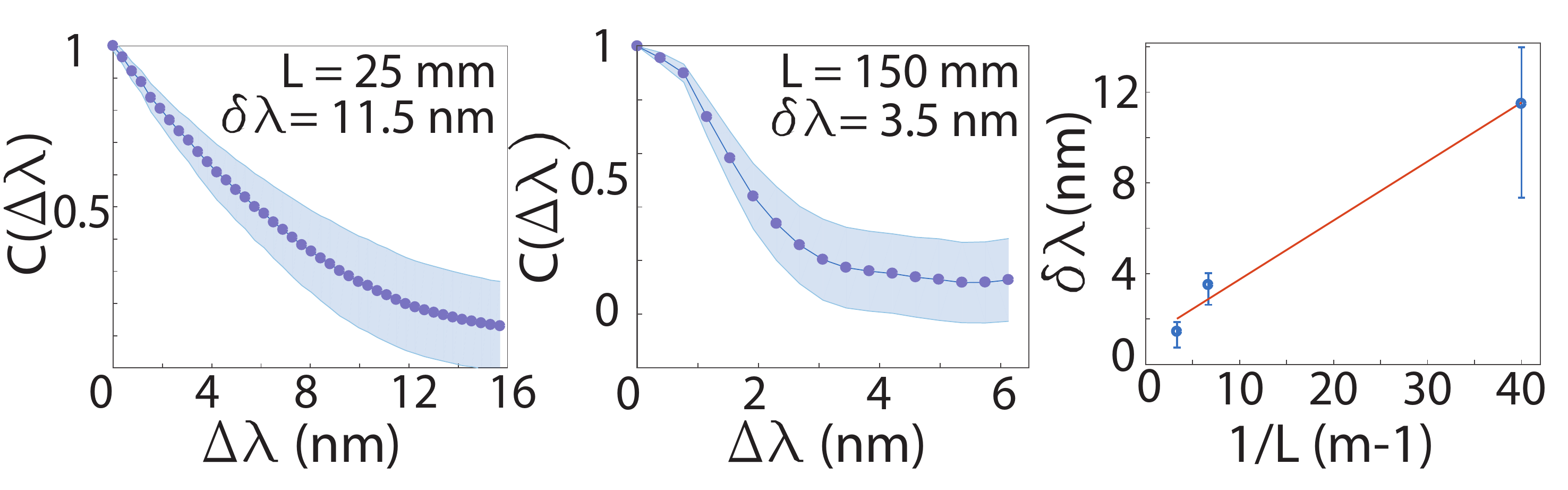}
		\caption{Fiber length-dependent characterization. (a) Spectral correlation function for 25~mm-length of MCMMF with a fiber core diameter of 50~\textmu m and spectral correlation width of 11.5~nm. (b) Spectral correlation function for a 50~\textmu m fiber core diameter and 150~mm-length of MCMMF with a spectral correlation width of 3.5~nm. (c) The spectral correlation width scales linearly with 1/L, where L is the fiber length.}
		\label{fig:Figure1supp}
\end{figure*}
	It was shown by Redding et al. that the spectral resolution of a multimode fiber spectrometer is dependent on the length of the fiber \cite{Redding2012}. Here, we review the dependence of the spectral width of MCMMFs on their corresponding length. The spectral correlation functions of two MCMMFs, with length, L, equal to 25~mm and 150~mm, respectively, are shown in Fig.~\ref{fig:Figure1supp}(a) and (b). Taking into account the spectral correlation width of the 30~cm-length of MCMMF which was used in the main article, Fig.~\ref{fig:Figure1supp}(c) shows a linear relationship between the spectral correlation width and 1/L. Hence, the spectral resolution of the MCMMF, when implemented in the spectral imaging system, will scale with fiber length. Ultimately, a higher spectral resolution can be achieved using a longer MCMMF due to a higher probability of modes mixing within a greater propagation length. 
	
	\section{DBSCAN}
	Density-based spatial clustering of applications with noise (DBSCAN) is a data clustering algorithm which can be used to detect over-densities of information in an image \cite{Ester1996}. When determining the coordinates of 10$^{3}$ speckle patterns produced by a MCMMF, we need to find a fast and efficient method to avoid software bottlenecks. DBSCAN allows us to detect objects in our images, before we can identify the coordinates of their corresponding locations. For the MCMMF output, the clusters correspond to speckle patterns produced by each fiber core, i.e. a cluster of high intensity speckles, with areas of low intensity pixels between each core. DBSCAN requires the following initial parameters: the maximum distance between individual speckles for them to be considered part of the same speckle pattern, incorporated into a parameter \textit{eps} or $\epsilon$, and the number of speckles contained within one speckle pattern, \textit{MinPts}. Unlike other well-known clustering algorithms, DBSCAN does not require prior knowledge of the total number of unique `clusters', or speckle patterns. This allows the algorithm to identify over a thousand speckle patterns produced by the MCMMF, as well as store the center coordinate information of each fiber core, within a couple of seconds. To minimize the detection of false clusters - for example, one speckle pattern produced by a single fiber core is identified as containing two or more clusters - the standard deviation of the distance between consecutive cluster centroids is determined, and a threshold is enforced to ensure that we only retain one coordinate for every speckle pattern. For the purposes of this study we used parameters of \textit{eps}~$= 3$ and \textit{MinPts}~$=13$ for the 30~cm MCMMF spectral imaging system. From the 1400 individual speckle patterns detected, we select an area of interest in the center of the MCMMF containing approximately 800 speckle patterns coordinates. The coordinates are used to identify the locations where pixel information should be collected and stored in the spectral intensity transmission matrix during calibration, and also post-calibration when using the spectral imaging system for detection purposes.
	
	\section{Spatial reconstructions across multiple spectral channels}
\begin{figure*}
		\centering
		\includegraphics[width=\linewidth]{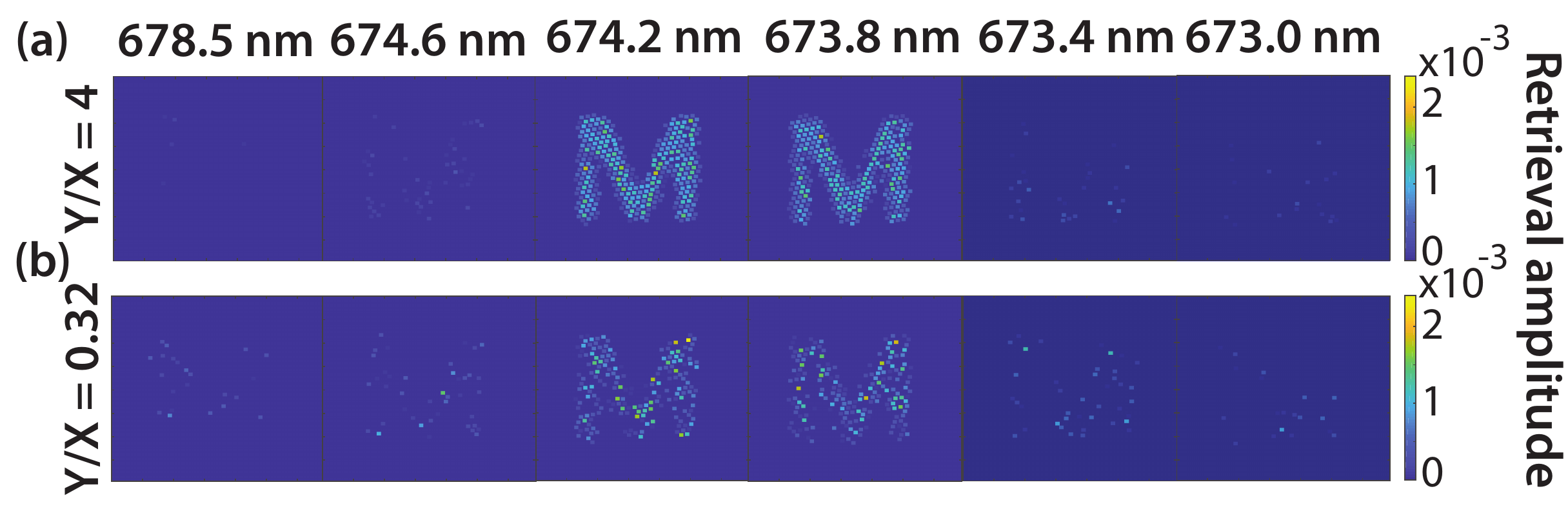}
		\caption{Spatial reconstruction across multiple spectral channels for different sampling rates: (a) when N$_{\lambda}$=1 and Y/X=4, the letter `M' is reconstructed in the spectral channels corresponding to 674.2~nm and 673.8~nm. Input information is not reconstructed in neighboring channels, however small noise fluctuations are observed. (b) when N$_{\lambda}$=1 and Y/X=0.32, the letter `M' is reconstructed in the same spectral bands as in (a), although larger amounts of noise are observed in neighboring spectral bands. }
		\label{fig:Figure2supp}
\end{figure*}
	
	An experiment was carried out to test the fidelity of the spectral imaging system across all calibrated spectral bands. The letter `M' was imaged using a central wavelength of 674.1~nm, and a STM with spectral channels calibrated in increments of 0.4~nm, was used to reconstruct the input signal. The reconstructions were carried out in the case of oversampling and under-sampling where the sampling rate, i.e. the ratio of the number of AOI camera pixels Y to the number of independent spectral components X, corresponded to Y/X=4 and Y/X=0.32, respectively, as shown in Fig.~\ref{fig:Figure2supp}. The resulting reconstructions were observed in 6 different spectral channels: $\lambda$~=~678.5~nm, 674.6~nm, 674.2~nm, 673.8~nm, 673.4~nm, 673.0~nm. When oversampling, as in Fig.~\ref{fig:Figure2supp}(a), the input signal is correctly reconstructed in the spectral bands corresponding to $\lambda$=674.2~nm and $\lambda$=673.8~nm. Apart from some small noise values, these spatial reconstructions are not observed in neighboring spectral channels. When under-sampling the input signal, as in Fig.~\ref{fig:Figure2supp}(b), the spatial reconstructions are detected in the correct spectral channels, however, we observe a larger amount of reconstruction noise in other spectral bands. This result shows that oversampling leads to high fidelity reconstruction of spatial and spectral information. Furthermore, under-sampling allows fast acquisition and minimal data collection, but with the compromise of more noisy reconstructions.

	\section{Angle dependence of incoming light}
	
\begin{figure*}
		\centering
		\includegraphics[width=\linewidth]{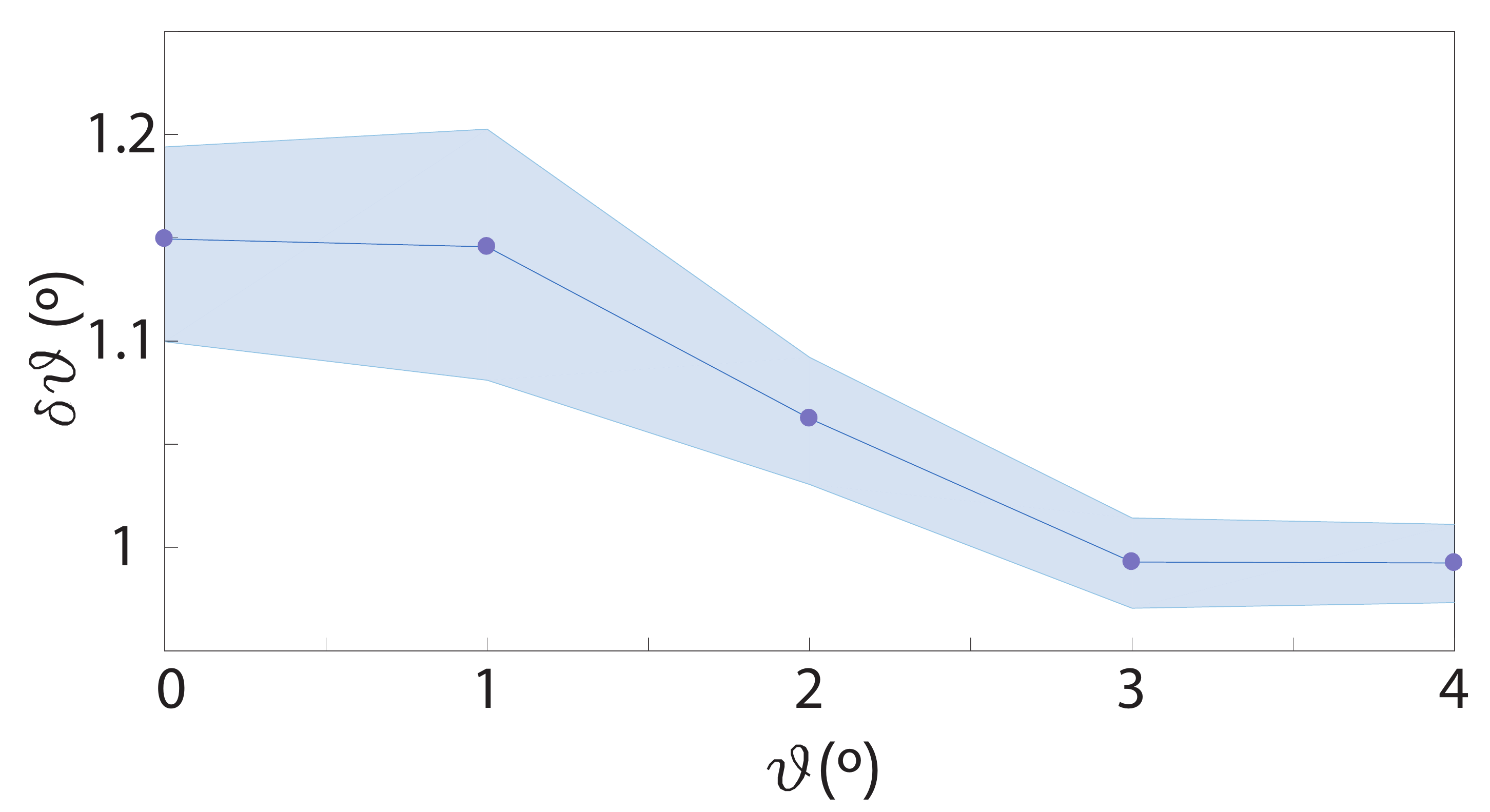}
		\caption{Angle dependence of speckle patterns. The relationship between angle correlation width and the incident angle of light.}
		\label{fig:Figure3supp}
\end{figure*}

	Further to details presented in Section~\ref{sec:level2}, the relationship between the incident angle of light propagating through the MCMMF and the resulting angle correlation width were determined. The angle correlation function was plotted for angles of incidence, $\theta$=0\textdegree, 1\textdegree, 2\textdegree, 3\textdegree, 4\textdegree, and the full width at half maximum was determined. The resulting angle correlation width was found to decrease as the incident angle was increased, as shown in Fig.~\ref{fig:Figure3supp}. This suggests that the aperture of the spectral imaging system is reduced as the number of speckle grains within each speckle pattern increases. 
	
	\section*{Acknowledgments}
	The authors would like to thank Jonathan Woods for the use of the CW External Cavity Diode Laser and help with its operation. Thanks also go to Southampton Astronomy Group for the introduction to DBSCAN, and to Laurent Daudet for help with compressive sensing.


\begin{thebibliography}{10}
	\newcommand{\enquote}[1]{``#1''}
	
	\bibitem{Hagen2013}
	N.~A. Hagen and M.~W. Kudenov, \enquote{Review of snapshot spectral imaging
		technologies,} {{Opt. Eng}} \textbf{52}, 090901 (2013).
	
	\bibitem{Bacon1988}
	R.~Bacon, G.~Adam, A.~Baranne, G.~Court{\`e}s, D.~Dubet, J.-P. Dubois,
	Y.~Georgelin, G.~Monnet, E.~Pecontal, and J.~Urios, \enquote{{The integral
			field spectrograph TIGER},} in \ {European Southern Observatory Conference
		and Workshop Proceedings}, (ESO, 1988)  \textbf{30}, pp. 1185.
	
	\bibitem{Peters2012}
	M.~A. Peters, T.~Groff, N.~J. Kasdin, M.~W. McElwain, M.~Galvin, M.~A. Carr,
	R.~Lupton, J.~E. Gunn, G.~Knapp, Q.~Gong, A.~Carlotti, T.~Brandt, M.~Janson,
	O.~Guyon, F.~Martinache, M.~Hayashi, and N.~Takato, \enquote{Conceptual
		design of the coronagraphic high angular resolution imaging spectrograph
		({CHARIS}) for the subaru telescope,} {{Proc. SPIE}}
	\textbf{8446}, 84467U (2012).
	
	\bibitem{Dwight2017}
	J.~G. Dwight and T.~S. Tkaczyk, \enquote{Lenslet array tunable snapshot imaging
		spectrometer ({LATIS}) for hyperspectral fluorescence microscopy,}
	{{Biomed. Opt. Express}} \textbf{8}, 1950--1964 (2017).
	
	\bibitem{Drory2015}
	N.~Drory, N.~MacDonald, M.~A. Bershady, K.~Bundy, J.~Gunn, D.~R. Law, M.~Smith,
	R.~Stoll, C.~A. Tremonti, D.~A. Wake, R.~Yan, A.~M. Weijmans, N.~Byler,
	B.~Cherinka, F.~Cope, A.~Eigenbrot, P.~Harding, D.~Holder, J.~Huehnerhoff,
	K.~Jaehnig, T.~C. Jansen, M.~Klaene, A.~M. Paat, J.~Percival, and C.~Sayres,
	\enquote{The {M}a{NGA} integral field unit fiber feed system for the sloan
		2.5 m telescope,} {{The Astronomical Journal}}
	\textbf{149}, 77 (2015).
	
	\bibitem{Eisenhauer2003}
	F.~Eisenhauer, K.~B. R.~Abuter, F.~Biancat-Marchet, H.~Bonnet, J.~Brynnel,
	R.~D. Conzelmann, B.~Delabre, R.~Donaldson, J.~Farinato, E.~Fedrigo,
	R.~Genzel, N.~N. Hubin, C.~Iserlohe, M.~E. Kasper, M.~Kissler-Patig, C.~R.
	G.~J.~Monnet, J.~Schreiber, S.~Stroebele, M.~Tecza, N.~A. Thatte, and
	H.~Weisz, \enquote{{SINFONI} - {I}ntegral field spectroscopy at 50
		milli-arcsecond resolution with the {ESO VLT},} {{Proc.
			SPIE}} \textbf{4841}, 1548 -- 1561 (2003).
	
	\bibitem{Kohlgraf2010}
	T.~W. Kohlgraf-Owens and A.~Dogariu, \enquote{Transmission matrices of random
		media: means for spectral polarimetric measurements,}
	{{Opt. Lett.}} \textbf{35}, 2236--2238 (2010).
	
	\bibitem{Xu2010}
	Z.~Xu, Z.~Wang, M.~E. Sullivan, D.~J. Brady, S.~H. Foulger, and A.~Adibi,
	\enquote{Multimodal multiplex spectroscopy using photonic crystals,}
	{{Opt. Express}} \textbf{11}, 2126--2133 (2003).
	
	\bibitem{Redding2012}
	B.~Redding and H.~Cao, \enquote{Using a multimode fiber as a high-resolution,
		low-loss spectrometer,} {{Opt. Lett.}} \textbf{37},
	3384--3386 (2012).
	
	\bibitem{Chakrabarti2015}
	M.~Chakrabarti, M.~L. Jakobsen, and S.~G. Hanson, \enquote{Speckle-based
		spectrometer,} {{Opt. Lett.}} \textbf{40}, 3264--3267
	(2015).
	
	\bibitem{Liew2016}
	S.~F. Liew, B.~Redding, M.~A. Choma, H.~D. Tagare, and H.~Cao,
	\enquote{Broadband multimode fiber spectrometer,} {{Opt.
			Lett.}} \textbf{41}, 2029--2032 (2016).
	
	\bibitem{Redding2013b}
	B.~Redding, S.~F. Liew, R.~Sarma, and H.~Cao, \enquote{{Compact spectrometer
			based on a disordered photonic chip},} {{Nat.
			Photonics}} \textbf{7}, 746--751 (2013).
	
	\bibitem{Mazilu2014}
	M.~Mazilu, T.~Vettenburg, A.~D. Falco, and K.~Dholakia, \enquote{Random
		super-prism wavelength meter,} {{Opt. Lett.}}
	\textbf{39}, 96--99 (2014).
	
	\bibitem{Hang2010}
	Q.~Hang, B.~Ung, I.~Syed, N.~Guo, and M.~Skorobogatiy, \enquote{Photonic
		bandgap fiber bundle spectrometer,} {{Appl. Opt.}}
	\textbf{49}, 4791--4800 (2010).
	
	\bibitem{Valley2016}
	G.~C. Valley, G.~A. Sefler, and T.~J. Shaw, \enquote{Multimode waveguide
		speckle patterns for compressive sensing,} {{Opt.
			Lett.}} \textbf{41}, 2529--2532 (2016).
	
	\bibitem{French2017}
	R.~French, S.~Gigan, and O.~L. Muskens, \enquote{Speckle-based hyperspectral
		imaging combining multiple scattering and compressive sensing in nanowire
		mats,} {{Opt. Lett.}} \textbf{42}, 1820--1823 (2017).
	
	\bibitem{GiganRotter2017}
	S.~Rotter and S.~Gigan, \enquote{Light fields in complex media: mesoscopic
		scattering meets wave control,} {{Rev. Mod. Phys.}}
	\textbf{89}, 015005 (2017).
	
	\bibitem{Popoff2010}
	S.~M. Popoff, G.~Lerosey, R.~Carminati, M.~Fink, A.~C. Boccara, and S.~Gigan,
	\enquote{Measuring the transmission matrix in optics: an approach to the
		study and control of light propagation in disordered media,}
	{{Phys. Rev. Lett.}} \textbf{104}, 100601 (2010).
	
	\bibitem{Liutkus2014}
	A.~Liutkus, D.~Martina, S.~Popoff, G.~Chardon, O.~Katz, G.~Lerosey, S.~Gigan,
	L.~Daudet, and I.~Carron, \enquote{Imaging with nature: compressive imaging
		using a multiply scattering medium,} {{Sci. Rep}}
	\textbf{4} (2014).
	
	\bibitem{Vellekoop2007}
	I.~M. Vellekoop and A.~P. Mosk, \enquote{Focusing coherent light through opaque
		strongly scattering media,} {{Opt. Lett.}} \textbf{32},
	2309--2311 (2007).
	
	\bibitem{Katz2012}
	O.~Katz, E.~Small, and Y.~Silberberg, \enquote{Looking around corners and
		through thin turbid layers in real time with scattered incoherent light,}
	{{Nat. Photon}} \textbf{6}, 549 EP -- (2012).
	
	\bibitem{Andreoli2015}
	D.~Andreoli, G.~Volpe, S.~Popoff, O.~Katz, S.~Gr\'{e}sillon, and S.~Gigan,
	\enquote{Deterministic control of broadband light through a multiply
		scattering medium via the multispectral transmission matrix,}
	{{Sci. Rep}} \textbf{5} (2015).
	
	\bibitem{Wang2015}
	P.~Wang and R.~Menon, \enquote{Ultra-high-sensitivity color imaging via a
		transparent diffractive-filter array and computational optics,}
	{{Optica}} \textbf{2}, 933--939 (2015).
	
	\bibitem{WangJOSA18}
	P.~Wang and R.~Menon, \enquote{Computational multispectral video imaging,}
	{{J. Opt. Soc. Am. A}} \textbf{35}, 189--199 (2018).
	
	\bibitem{Sahoo2017}
	S.~K. Sahoo, D.~Tang, and C.~Dang, \enquote{Single-shot multispectral imaging
		with a monochromatic camera,} {{Optica}} \textbf{4},
	1209--1213 (2017).
	
	\bibitem{Ploschner2015}
	M.~Ploschner, T.~Tyc, and T.~Cizmar, \enquote{Seeing through chaos in multimode
		fibres,} {{Nat. Photon}} \textbf{9} (2015).
	
	\bibitem{Porat2016}
	A.~Porat, E.~R. Andresen, H.~Rigneault, D.~Oron, S.~Gigan, and O.~Katz,
	\enquote{Widefield lensless imaging through a fiber bundle via speckle
		correlations,} {{Opt. Express}} \textbf{24},
	16835--16855 (2016).
	
	\bibitem{Redding2014}
	B.~Redding, M.~Alam, M.~Seifert, and H.~Cao, \enquote{{High-resolution and
			broadband all-fiber spectrometers},} {{Optica}}
	\textbf{1}, 175--180 (2014).
	
	\bibitem{Wan2015}
	N.~H. Wan, F.~Meng, T.~Schr{\"o}der, R.~Shiue, E.~H. Chen, and D.~Englund,
	\enquote{High-resolution optical spectroscopy using multimode interference in
		a compact tapered fibre,} {{Nat. Commun}} \textbf{6},
	7762 EP -- (2015).
	
	\bibitem{Wagadarikar2008}
	A.~Wagadarikar, R.~John, R.~Willett, and D.~Brady, \enquote{Single disperser
		design for coded aperture snapshot spectral imaging,}
	{{Appl. Opt.}} \textbf{47}, B44--B51 (2008).
	
	\bibitem{Gehm2007}
	M.~E. Gehm, R.~John, D.~J. Brady, R.~M. Willett, and T.~J. Schulz,
	\enquote{Single-shot compressive spectral imaging with a dual-disperser
		architecture,} {{Opt. Express}} \textbf{15},
	14013--14027 (2007).
	
	\bibitem{Cao2016}
	X.~Cao, T.~Yue, X.~Lin, S.~Lin, X.~Yuan, Q.~Dai, L.~Carin, and D.~J. Brady,
	\enquote{Computational snapshot multispectral cameras: toward dynamic capture
		of the spectral world,} {{IEEE Signal Process. Mag.}}
	\textbf{33}, 95--108 (2016).
	
	\bibitem{Willett2013}
	R.~W. Willett, M.~F. Duarte, M.~A. Davenport, and R.~G. Baranuik,
	\enquote{Sparsity and structure in hyperspectral imaging : sensing,
		reconstruction, and target detection,} {{IEEE Signal
			Process. Mag.}} \textbf{31}, 116--126 (2013).
	
	\bibitem{IAugust2013}
	I.~August, Y.~Oiknine, M.~AbuLeil, I.~Abdulhalim, and A.~Stern,
	\enquote{Miniature compressive ultra-spectral imaging system utilizing a
		single liquid crystal phase retarder,} {{Sci. Rep}}
	\textbf{6}, (2016).
	
	\bibitem{YAugust2013}
	Y.~August, C.~Vachman, Y.~Rivenson, and A.~Stern, \enquote{Compressive
		hyperspectral imaging by random separable projections in both the spatial and
		the spectral domains,} {{Appl. Opt.}} \textbf{52},
	D46--D54 (2013).
	
	\bibitem{Rueda2015}
	H.~Rueda, D.~Lau, and G.~R. Arce, \enquote{Multi-spectral compressive snapshot
		imaging using rgb image sensors,} {{Opt. Express}}
	\textbf{23}, 12207--12221 (2015).
	
	\bibitem{Fowler2014}
	J.~E. Fowler, \enquote{Compressive pushbroom and whiskbroom sensing for
		hyperspectral remote-sensing imaging,} in \ {Proceedings of IEEE
		International Conference on Image Processing (ICIP),}  (IEEE, 2014), pp. 684--688.
	
	\bibitem{Lin2014}
	X.~Lin, Y.~Liu, J.~Wu, and Q.~Dai, \enquote{Spatial-spectral encoded
		compressive hyperspectral imaging,} {{ACM Trans.
			Graph.}} \textbf{33}, 233:1--233:11 (2014).
	
	\bibitem{Candes2006b}
	E.~J. Candes, J.~Romberg, and T.~Tao, \enquote{Robust uncertainty principles:
		exact signal reconstruction from highly incomplete frequency information,}
	{{IEEE Trans. Inf. Theory}} \textbf{52}, 489--509
	(2006).
	
	\bibitem{Donoho2006}
	D.~L. Donoho, \enquote{Compressed sensing,} {{IEEE Trans.
			Inf. Theory}} \textbf{52}, 1289 -- 1306 (2006).
	
	\bibitem{duarte2008single}
	M.~F. Duarte, M.~A. Davenport, D.~Takhar, J.~N. Laska, T.~Sun, K.~E. Kelly,
	R.~G. Baraniuk, \enquote{Single-pixel imaging via compressive
		sampling,} {{IEEE Signal Process. Mag.}} \textbf{25}, 83
	(2008).
	
	\bibitem{Woods2016}
	J.~Woods, \enquote{A mode-locked diode laser frequency comb for ultracold
		atomic physics experiments,} Ph.D. thesis, University of Southampton (2016).
	
	\bibitem{Ester1996}
	M.~Ester, H.~Kriegel, J.~Sander, and X.~Xu, \enquote{A density-based algorithm
		for discovering clusters a density-based algorithm for discovering clusters
		in large spatial databases with noise,} in \ {Proceedings of the Second
		International Conference on Knowledge Discovery and Data Mining,}  (AAAI
	Press, 1996), KDD-96, pp. 226--231.
	
	\bibitem{CVX2013}
	M.~Grant and S.~Boyd, \enquote{{CVX}: {MATLAB} software for disciplined convex
		programming, version 2.0 beta,} http://cvxr.com/cvx (2013).
	
	\bibitem{Pan1994}
	K.~Pan, C.~Uang, F.~Cheng, and F.~T.~S. Yu, \enquote{Multimode fiber sensing by
		using mean-absolute speckle-intensity variation,}
	{{Appl. Opt.}} \textbf{33}, 2095--2098 (1994).
	
	\bibitem{Choi1997}
	H.~Choi, H.~F. Taylor, and C.~E. Lee, \enquote{High-performance fiber-optic
		temperature sensor using low-coherence interferometry,}
	{{Opt. Lett.}} \textbf{22}, 1814--1816 (1997).
	
	\bibitem{Liu2007}
	Y.~Liu and L.~Wei, \enquote{Low-cost high-sensitivity strain and temperature
		sensing using graded-index multimode fibers,} {{Appl.
			Opt.}} \textbf{46}, 2516--2519 (2007).
	
	\bibitem{Amitonova2015}
	L.~V. Amitonova, A.~P. Mosk, and P.~W.~H. Pinkse, \enquote{Rotational memory
		effect of a multimode fiber,} {{Opt. Express}}
	\textbf{23}, 20569--20575 (2015).
	
	\bibitem{Saleh2001}
	B.~E. Saleh and M.~C. Teich, \ {Fiber-Optic Communications in Fundamentals
		of Photonics} (Wiley-Blackwell, 2001), chap.~22, pp. 874--917.
	
\end{thebibliography}
\end{document}